\documentclass[a4paper,11pt]{article}

\usepackage{lipsum}
\usepackage{caption}
\usepackage{subcaption}

\pdfoutput=1 

\usepackage{jinstpub} 

\usepackage{lineno}
\usepackage{siunitx}

\title{First Measurements on the Timespot1 ASIC: a Fast-Timing, High-Rate Pixel-Matrix Front-End}

\author[a,b]{Lorenzo Piccolo,}
\author[c]{Sandro Cadeddu,}
\author[d,e]{Luca Frontini,}
\author[c]{Adriano Lai,}
\author[d,e]{Valentino Liberali,}
\author[a]{Angelo Rivetti,}
\author[d,e]{Alberto Stabile}

\affiliation[a]{INFN sezione di Torino,\\Via P. Giuria 1 Torino, Italy}
\affiliation[b]{Politecnico di Torino,\\Corso Duca degli Abruzzi 24 Torino, Italy}

\affiliation[c]{INFN sezione di Cagliari,\\S.P. per Sestu km 1.0 Monserrato (Cagliari), Italy}

\affiliation[d]{INFN sezione di Milano,\\Via Celoria 16 Milano, Italy}
\affiliation[e]{Università degli Studi di Milano - Dipartimento di Fisica,\\Via Celoria 16 Milano, Italy}

\emailAdd{lorenzo.piccolo@to.infn.it}

\abstract{

This work presents the first measurements performed on the Timespot1 ASIC.    As the second prototype developed for the TimeSPOT project, the ASIC features a \SI{32x32}{} channels hybrid-pixel matrix. Targeted to space-time tracking applications in High Energy Physics experiments, the system aims to achieve a time resolution of \SI{30}{ps} or better at a maximum event rate of  \SI{3}{MHz\per channel} with a Data Driven interface.   Power consumption can be programmed to range between \SI{1.2}{W\per cm^{2}} and \SI{2.6}{W\per cm^2}.   The presented results include a description of the ASIC operation and a first characterization of its performance in terms of time resolution.

}

\keywords{Hybrid detectors, Particle tracking detectors, Timing detectors, Front-end electronics for detector readout, VLSI circuits, Analogue electronic circuits, Digital electronic circuits.}

\proceeding{TWEPP2021: Topical Workshop on Electronics for Particle Physics \\
  20-24 September 2021\\
  Online event}

\begin{document}
\maketitle
\flushbottom

\section{Introduction}

Future upgrades on High Energy Physics experiments aim to improve their capability to detect rare events by increasing the beam luminosity \cite{hllhc}. When operating in high luminosity regimes, current tracking techniques will no longer be sufficient to efficiently reconstruct the event. A proposed solution to this problem is adding a fine time measurement to the position information \cite{lhcbu2}\cite{cmsu2}\cite{atlasu2}.
The TimeSPOT project \cite{tspotweb} aims to build a small scale telescope demonstrator suitable for future experiments.   The activity of the projects consists in both designing and testing of the whole detector including its sensors, front-end ASIC and readout electronics.

This article presents the first results from electrical tests on the Timespot1 ASIC. This front-end chip was designed to cope with a required timing resolution of \SI{50}{ps} per single hit with an event rate per unit area larger than \SI{11.6}{\giga \hertz \per  \centi \meter^{2}}.    
 These requirements must also be met while keeping the power consumption per unit area below \SI{1.5}{\watt \per \centi \meter ^2} in order to be compatible with cooling.    Furthermore, the candidate TimeSPOT 3D-Silicon sensor has experimentally proven to be capable of reaching a intrinsic time resolution better than \SI{20}{ps} \cite{si3d1}\cite{si3d2}, establishing a new challenge for the FE electronics.  
In section \ref{sec:arc} the ASIC architecture is briefly described.
Pixel performance measurements are illustrated from the point of view of the time resolution of both the Time to Digital converter (TDC) in section \ref{sec:tdc} and the Analog Front-End electronics (AFE) in section \ref{sec:afe}. 

\section{Chip Architecture}\label{sec:arc}

\begin{figure}[htbp]
    \centering
        \begin{subfigure}[t]{0.44\textwidth}
                \centering
                \includegraphics[width=0.85\textwidth]{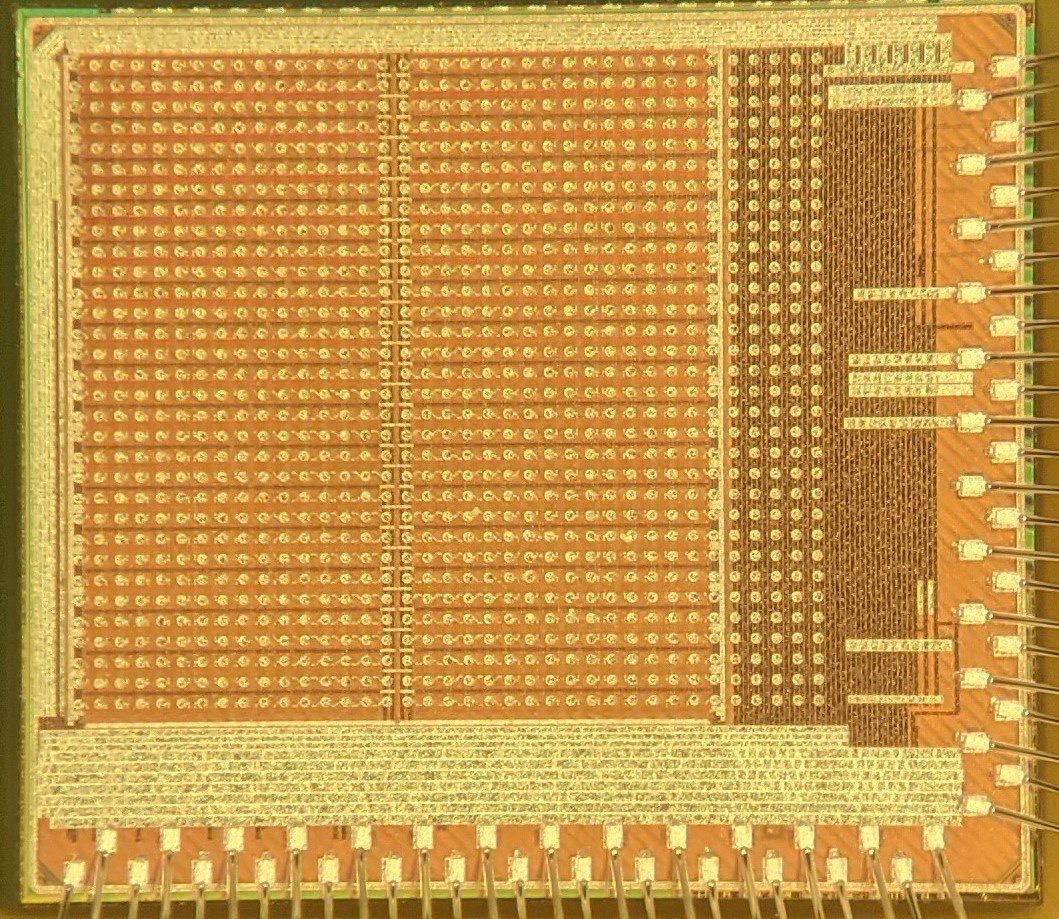}
                \caption{Timespot1 ASIC.\\ Chip size: \SI{2.6 x 2.3}{\milli \meter}.  
                }\label{fig:asic} 
         \end{subfigure}
         \hfill
         \begin{subfigure}[t]{0.54\textwidth}
                \centering
                \includegraphics[width=\textwidth,]{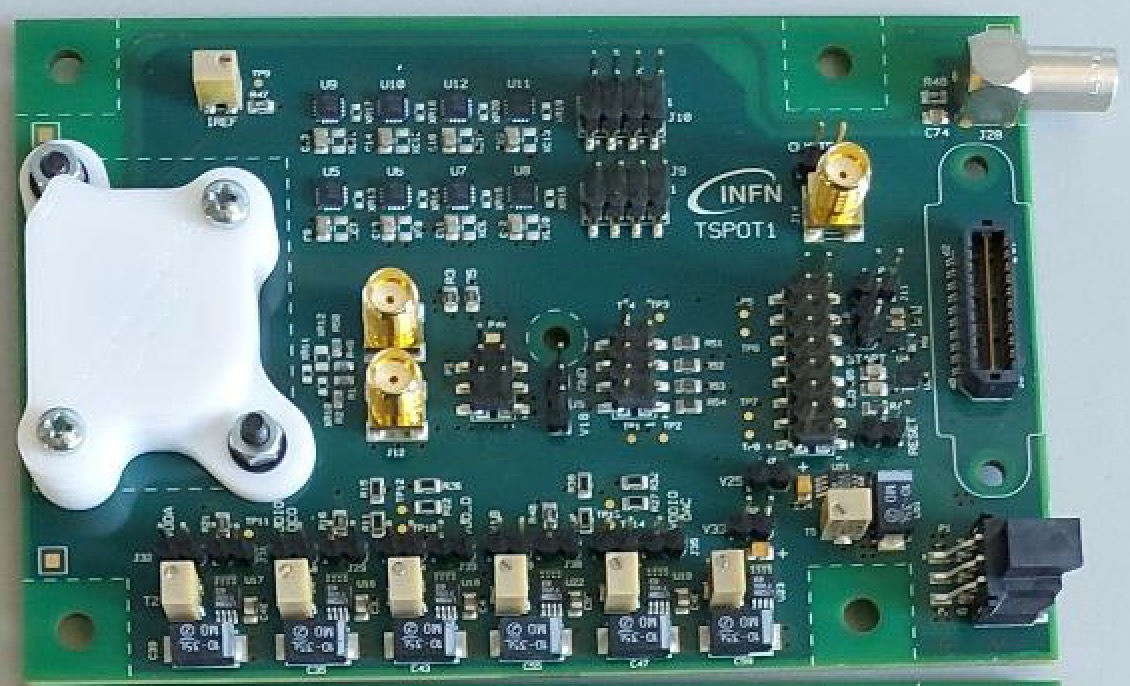}
                \caption{TSPOT1 PCB. Board size: \SI{8x12}{cm}. The ASIC is mounted on the left side under the white removable cover. 
                }\label{fig:pcb}
        \end{subfigure}
        \caption{}\label{}
\end{figure}

A picture of Timespot1 is shown in figure \ref{fig:asic}.  The \SI{2.6 x 2.3}{\milli \meter}  chip is manufactured in a \SI{28}{nm} CMOS commercial technology.  This prototype is bump-bondable to sensors with a \SI{32x32}{} pixel matrix with a pixel pitch of \SI{55}{\micro \meter}. Five more columns of 32 dummy pixels are inserted to ensure mechanical stability.   Input-Output signals and supply voltages are delivered through wire-bonding. The wire-bond pads are located on two adjacent sides of the chip making it two-side tileable.  The ASIC has a data-driven interface.

The pixel matrix is organized in two symmetrical blocks of \SI{16x32}{} pixels. Each block includes two service columns: a digital one for pixel generated data distribution and pixel programming, and an analog one incorporating four independent service DACs, a band-gap and a programmable cell used to perform a fine setting of power consumption of the AFE components.   Analog and digital circuits have independent power and ground nets in order to prevent cross talk, these nets are also included in the respective columns. For the same reason all the analog circuit has been realized inside dedicated triple-n-wells.  All the nets are then redistributed by a repeated double row configuration of \SI{16x2}{} pixels.    Each pixel has a reduced pitch of \SI{50}{\micro \meter} in the horizontal direction compared to the bond-pad matrix.    In this way every 16 pixel \SI{75}{\micro \meter} can be reserved to host the lateral service columns, making the design indefinitely repeatable.

\begin{figure}[htbp]
\centering 
\includegraphics[width=\textwidth]{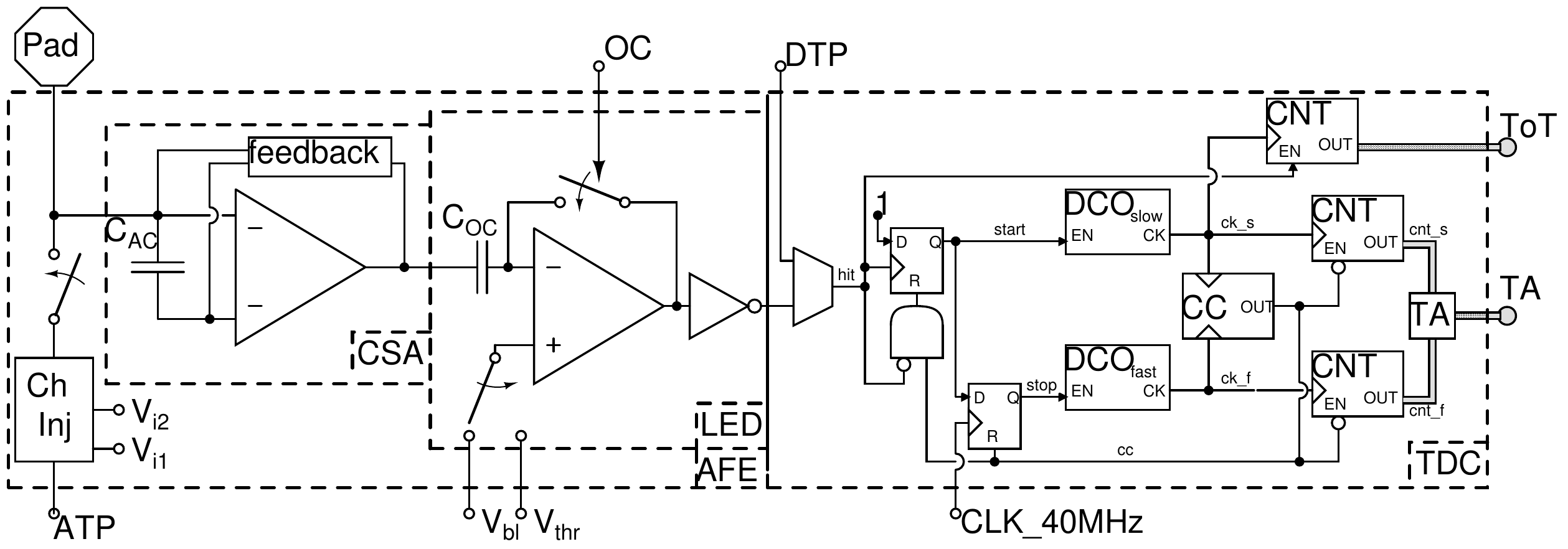}
\caption{\label{fig:schem} Schematic representation of the pixel architecture.}
\end{figure}

The pixel architecture is presented in figure \ref{fig:schem}. Every pixel includes the AFE directly connected to the sensor pad as well as its dedicated TDC. The AFE chain is comprised of an input and inverter based Charge Sensitive Amplifier (CSA) with DC current compensation and a Leading Edge Discriminator (LED) with discrete-time Offset Compensation (OC).    The TDC is based on a Vernier architecture with its two Digital Controlled Oscillators (DCO) clocked around \SI{1}{\giga \hertz}.   Every channel generates a 24 bit word which is then transmitted serially at \SI{160}{\mega \hertz}.   A charge injection capacitance is also included in every channel for testing purpose.

Data generated at pixel level are then redistributed to the chip periphery using four independent Read Out Trees (ROT).  Each of these combinational blocks connects one fourth of the matrix to two multiplexed output links after a proper de-randomization by mean of a FIFO layer.    The ASIC has in total 8 LVDS output drivers at \SI{1.28}{Gb/s} each.
Configuration is provided by an I$^2$C interface.   Additionally an LVDS receiver is used to provide the system clock and a CMOS input is used as a Start signal providing absolute time reference.   

The TSPOT1 PCB (in figure \ref{fig:pcb}) was designed both for chip standalone testing and as part of the final demonstrator.  It provides ASIC grounding and power supplies using on-board LDOs as well as sensor biasing.  The board also interfaces the ASIC with FPGA via QTH connector for data IO and provides the system clock via SMA connectors.

\section{TDC Measurements}\label{sec:tdc}

The TDC measures the phase between the input signal and the \SI{40}{\mega \hertz} reference clock with a resolution dependent on the frequency difference of its two DCOs. The input signal triggers the activation of the slower DCO while the next \SI{40}{\mega \hertz} clock rising edge, providing the stop signal, activates the faster DCO.    Every period the phase between the two oscillators shrinks until it reverts. The count of the number of periods when this condition arises encodes the timing measurement. This kind of measurement will be referred to as Time of Arrival (TA).   The TDC will simultaneously measure the Time over Threshold (ToT) of a signal.  Due to the higher jitter associated with this parameter, a lower resolution is required for this measure: it is performed by directly counting  the number of DCO period between the rising and falling edges of the pulse.   DCOs calibration is crucial to extrapolate a reliable measure.   This calibration is automatically operated by a per pixel self procedure. 

\begin{figure}[htbp]
    \centering
        \begin{subfigure}[t]{0.49\textwidth}
                \centering
                    \includegraphics[width=\textwidth]{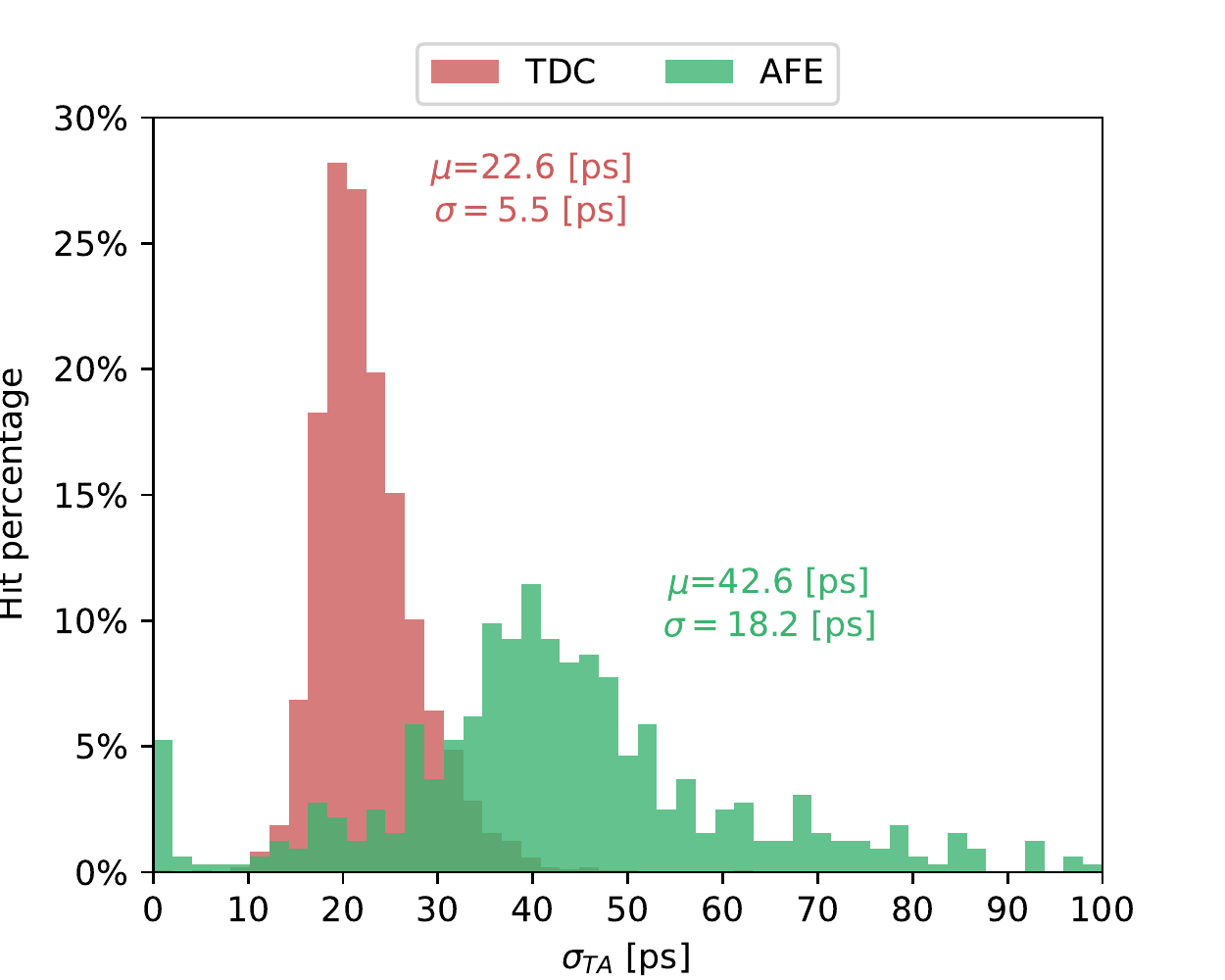}
                    \caption{Red histograms: $\sigma_{TA}$ on 100 repeated DTP across 1024 channels and 7 phases. Green histogram: $\sigma_{TA}$ on 100 repeated ATP across 512 channels for a \SI{2}{fC} input signal (MIP), the TDC contribution has been square subtracted.}\label{fig:hist} 
         \end{subfigure}
         \hfill
         \begin{subfigure}[t]{0.49\textwidth}
                \centering
                    \includegraphics[width=\textwidth,]{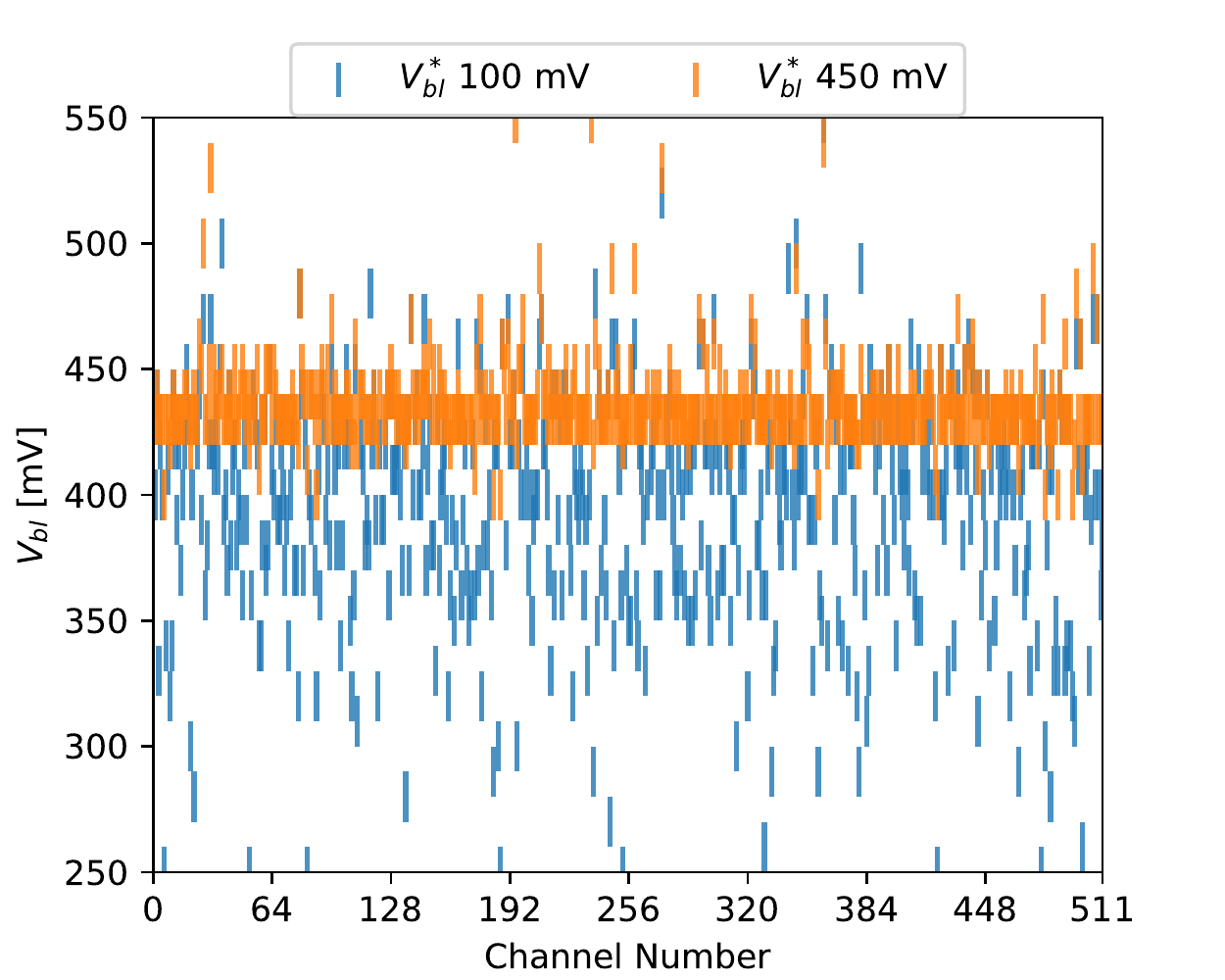}
                    \caption{Baseline values obtained from threshold scan on 512 channels. The desired baseline is $V_{bl}^*$. At \SI{100}{mV} OC fails to bringing $V_{bl}$ to uncompensated higher values. At \SI{450}{mV} the OC is working properly for most of channels.}\label{fig:OC}
        \end{subfigure}
        \caption{}\label{}
\end{figure}

From the point of view of the self-test capability a Digital Test Pulse (DTP) can be injected.  The signal can be programmed by changing its phase to 7 different sub-reference values and its width to 32 values.    In order to measure the timing resolution the same measure has been repeated multiple times, the standard deviation on this measure is then used to quantify the resolution.    This analysis can be repeated for different parameters in order to study dependencies.    This communication focuses on measurement on TA resolution since its constitutes the most critical measurement for timing.    The ToT measurement has exhibited an overall time resolution  of \SI{0.6}{ns} which is adequate to measure the intended signal.  TA measurements have been repeated 100 times for each channel and for all the 7 input phases. Standard deviation of TA ($\sigma_{TA}$) is computed for each case, the results are collected in the histogram figure \ref{fig:hist}.    In this condition the TDC consumes \SI{25}{\micro \watt} of power.

\section{Analog FE Measurements}\label{sec:afe}

The AFE adapts the sensor current signal into a digital pulse to be processed by the TDC.  The CSA produces a steep voltage signal with amplitude proportional to the input integrated charge. This charge is collected on the parasitic feedback capacitance and discharged with a constant current. In this way the signal ToT is proportional to the input charge enabling its measurement.    Corrections based on ToT measurements can be used to reduce the effect of the time walk.    LED offset compensation is operated by firstly saving the desired baseline voltage $V_{bl}^*$ on the memory capacitance $C_{OC}$ and then rising the threshold to $V_{thr}$.    This operation is performed by switching between two voltages provided by dedicated DACs. 

The AFE can be tested by injecting an Analog Test Pulse (ATP) by switching between two voltages.    In this way  a charge up to \SI{7}{\femto \coulomb} can be injected.    The ATP is always injected synchronously with the next reference clock rising edge, its TA represents the systematic propagation delay of the AFE.  The signal is then directly measured by the pixel TDC.  CSA signals can be characterized by threshold reconstruction on repeated signals.  
In order to quantify AFE contribution to the total $\sigma_{TA}$, the TDC contribution can be square subtracted from it.    The total is computed from ATP, while the TDC contribution from DTP.

\begin{figure}[htbp]
\centering 
\includegraphics[width=\textwidth]{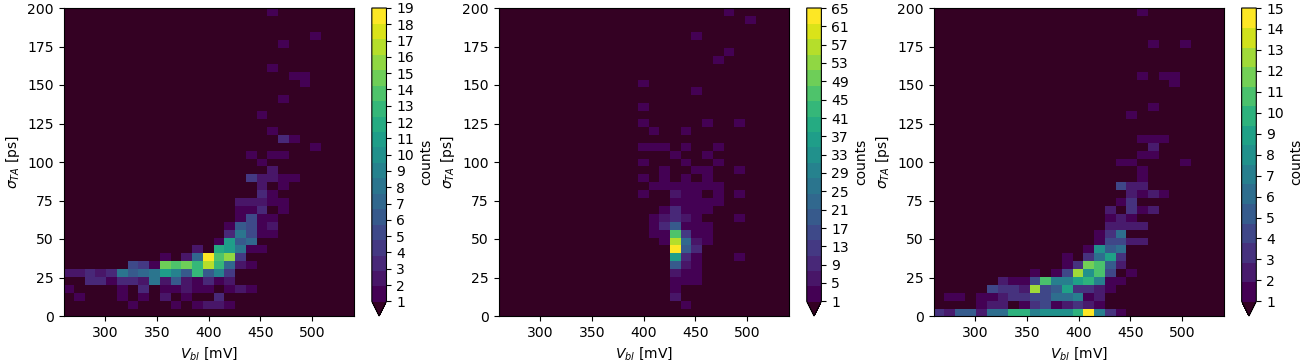}
\caption{\label{fig:blVSta} Correlation between measured $\sigma_{TA}$ and  $V_{bl}$. From left to right: $V_{bl}^*$ \SI{100}{mV} (case of OC failure), OC at $V_{bl}^*$ \SI{450}{mV} and $V_{bl}^*$ \SI{100}{mV} with TDC contribution removed.    AFE total resolution is affected by LED baseline position. Channels with low $V_{bl}$ shows CSA intrinsic resolution. }
\end{figure}

The AFE resolution is presented in figure \ref{fig:hist}.  In this condition the circuit consumes \SI{15}{\micro \watt} of power. An issue with OC was found: the circuit is unable to set the base line to low values.  This behaviour can be attributed to an unexpected voltage value across $C_{OC}$ before compensation.  The default voltage of this node is closer to $V_{DD}$ compared to the one indicated by simulation, making the compensation time insufficient to move actual $V_{bl}$ to the lower values.   This behaviour is presented in figure \ref{fig:OC}.    The OC issue forces the setting of $V_{bl}^*$ to \SI{450}{mV} ($V_{DD}/2$) or higher.    In this regime the P-type input differential cell  of LDE limits its bias currents resulting in a loss of bandwidth and therefore slew-rate.    By correlating $V_{bl}$ position with $\sigma_{TA}$ it is possible to understand this behaviour as presented in the plots of  figure \ref{fig:blVSta}.    This analysis shows that the actual CSA performance is masked by the LED issue and the TDC resolution. The CSA is capable to produce signal with a timing resolution better than \SI{20}{ps}.  It is noted that the OC compensation issue is not an intrinsic problem of the LDE design and therefore it can be solved with minor adjustments on the scheme. 

\section{Conclusions}
The Timespot1 ASIC has been tested in standalone configuration.
The TDC resolution is below \SI{50}{ps}, with an average of \SI{23}{ps}.   From the point of view of the AFE the resolution has been quantified to be under \SI{100}{ps} with an average of \SI{43}{ps}.    All measures have been performed within the specified power consumption constraint of \SI{40}{\micro \watt} per pixel.
The tests illustrated in the present paper show the possibility of improving the performance of the proposed architecture with minor corrections.
Measurements with the actual sensor matrix and particle generated signals will be performed in the near future.


\acknowledgments

This work was supported by the Fifth Scientific Commission (CSN5) of the Italian National Institute
for Nuclear Physics (INFN), within the Project TimeSPOT and by the ATTRACT-INSTANT-P1 (EC GA 777222) INSTANT project.

\end{document}